\documentclass[conference]{IEEEtran}
\IEEEoverridecommandlockouts

\usepackage{amsmath,epsfig}
\usepackage{multirow}
\usepackage{array}
\usepackage{booktabs}
\usepackage{tabu}
\usepackage{threeparttable}
\usepackage{url}

\begin{document}

\title{Lossless Point Cloud Attribute Compression with Normal-based Intra Prediction\\
\thanks{{*}Jianwen Chen is the corresponding author. © 2021 IEEE. Personal use of this material is permitted. Permission from IEEE must be obtained for all other uses, in any current or future media, including reprinting/republishing this material for advertising or promotional purposes, creating new collective works, for resale or redistribution to servers or lists, or reuse of any copyrighted component of this work in other works.}
}

\author{\IEEEauthorblockN{Qian Yin, Qingshan Ren, Lili Zhao, Wenyi Wang, and Jianwen Chen{*}}
\IEEEauthorblockA{\textit{School of Information and Communication Engineering} \\
\textit{University of Electronic Science and Technology of China}\\
Chengdu, China \\
chenjianwen@uestc.edu.cn}
}

\maketitle

\begin{abstract}
The sparse LiDAR point clouds become more and more popular in various applications, e.g., the autonomous driving. However, for this type of data, there exists much under-explored space in the corresponding compression framework proposed by MPEG, i.e., geometry-based point cloud compression (G-PCC). In G-PCC, only the distance-based similarity is considered in the intra prediction for the attribute compression. In this paper, we propose a normal-based intra prediction scheme, which provides a more efficient lossless attribute compression by introducing the normals of point clouds. The angle between normals is used to further explore accurate local similarity, which optimizes the selection of predictors. We implement our method into the G-PCC reference software. Experimental results over LiDAR acquired datasets demonstrate that our proposed method is able to deliver better compression performance than the G-PCC anchor, with $2.1\%$ gains on average for lossless attribute coding.
\end{abstract}


\section{Introduction}
\label{sec:intro}
With rapid development of 3D sensing and capturing technologies, point clouds, which have the capacity of representing spatial structures and surface properties of 3D objects or scenes, are often encountered in various fields, e.g., the autonomous driving, the heritage reconstruction and 3D immersive communication \cite{apps}. However, it is well-known that point clouds have the unorganized distribution in 3D space and consist of millions of points, which imposes the burden on the limited transmission bandwidth and storage space. Therefore, the compression of point clouds is indispensable but challenging. Therefore, it is necessary to explore more effective point cloud compression (PCC) schemes.

\begin{figure}[t]
\begin{minipage}[b]{0.31\linewidth}
  \centering
 \centerline{\epsfig{figure=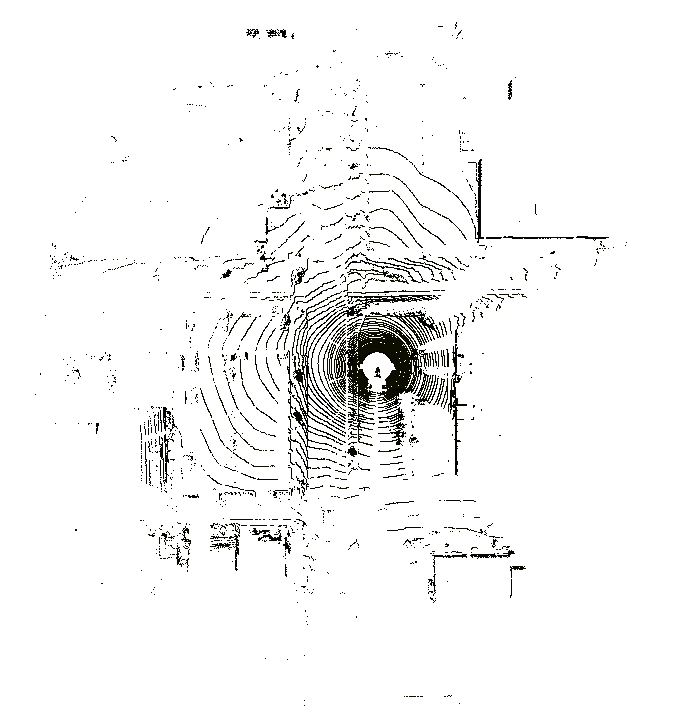, width=3.2cm}}
  \centerline{(a)}\medskip
\end{minipage}
\begin{minipage}[b]{0.32\linewidth}
  \centering
 \centerline{\epsfig{figure=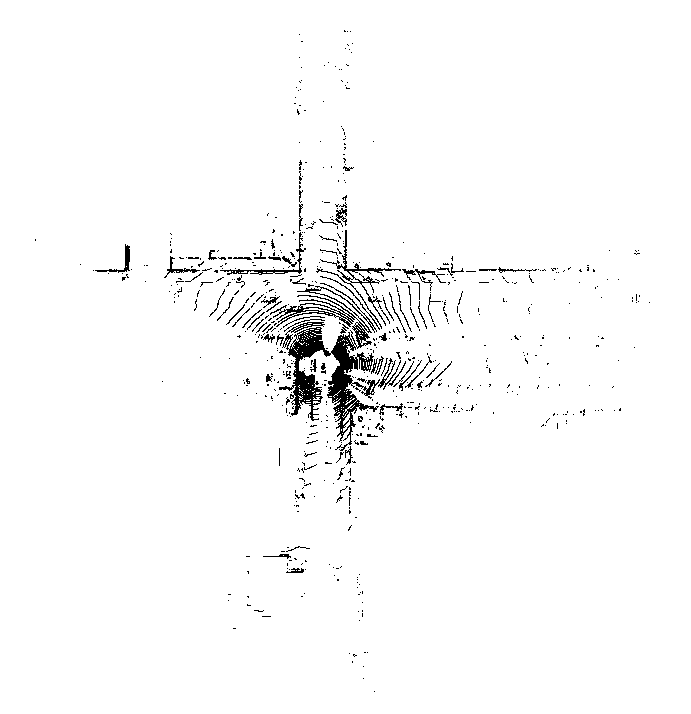,width=3.7cm}}
  \centerline{(b)}\medskip
\end{minipage}
\hfill
\begin{minipage}[b]{0.33\linewidth}
  \centering
 \centerline{\epsfig{figure=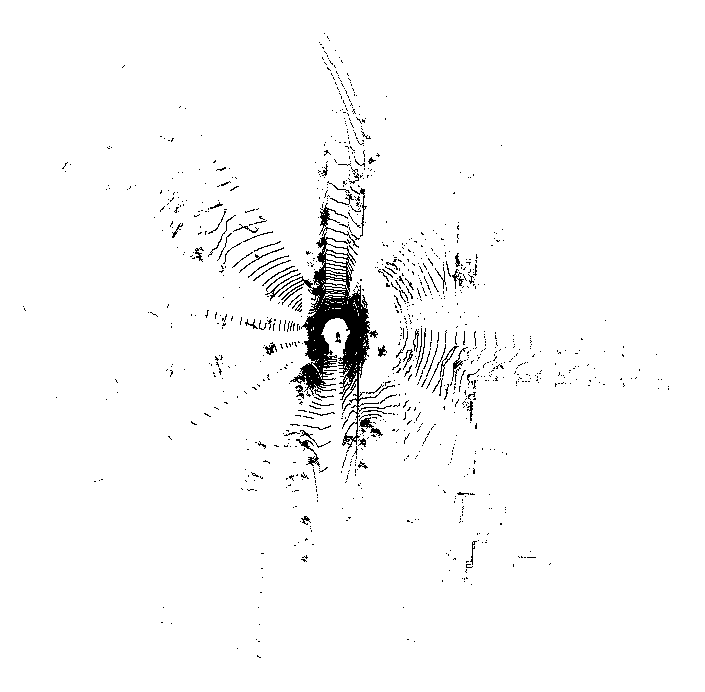,width=3.5cm}}
  \centerline{(c)}\medskip
\end{minipage}
\caption{Three single frame snapshots with bird views from Ford sequences provided by MPEG, which are point clouds acquired by LiDAR sensors. (a) Ford\_02\_vox1mm-0100. (b) Ford\_01\_vox1mm-0100. (c) Ford\_03\_vox1mm-0200.}
\label{fig:res} 
\end{figure}

Due to the potential needs of point cloud related applications, the Moving Picture Experts Group (MPEG) establishes the MPEG-3DG subgroup to exploit universal PCC frameworks. Two standardized test models are proposed: the video-based PCC (V-PCC) \cite{VPCC} and the geometry-based PCC (G-PCC) \cite{GPCC}. For V-PCC, 3D point clouds are projected into the 2D domain, and then coded by using the existing video codec (e.g., High Efficiency Video Coding, HEVC~\cite{HEVC}), which is more suitable for the dense point cloud compression. The G-PCC, by contrast, performs better for the sparse point cloud by compressing point clouds in the original 3D space. It is noted that point clouds typically consist of geometry information (i.e., 3D coordinates) and attribute information (e.g., colors, reflectances and normals). In the existing PCC schemes, the geometry and attribute information are coded separately. In this work, we focus on the attribute lossless coding in G-PCC for point clouds acquired by LiDAR sensors. Some examples can be seen in Fig. $1$.

Among the existing methods for the point cloud attribute coding, one popular strategy is the \emph{Graph Fourier Transform} (GFT) (e.g., \cite{GFT1}, \cite{GFT2}, \cite{GFT3}). In \cite{GFT1}, Zhang \textit{et al}. proposed an attribute compression method based on the graph transform. This scheme delivers a more effective PCC by using GFT instead of the traditional DCT, but generates many isolated sub-graphs when it comes to the sparse point clouds. To address this issue, Robert \textit{et al}. used the k-nearest neighbours (KNNs) method in \cite{GFT2} and Shao \textit{et al}. introduced Laplacian sparsity in \cite{GFT3} to optimize the graph transform respectively. However, due to much higher computational complexity introduced by eigenvalue decompositions, it is difficult for these graph-based methods to achieve real-time PCC. 

\begin{figure}[htb]
  \centering
  \includegraphics[width=0.35\textwidth, height=0.38\textheight]{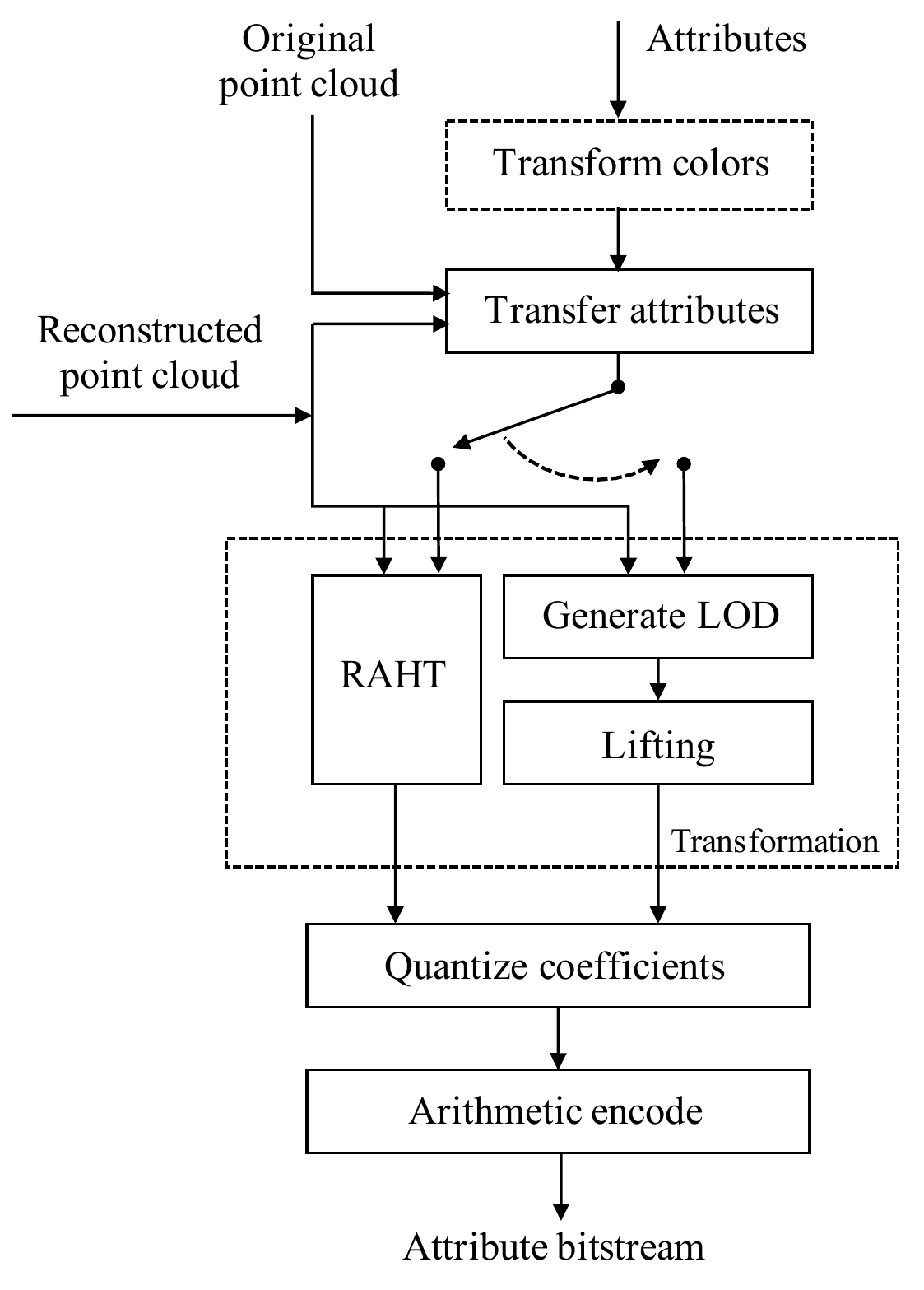}
  \caption{The overview of the attribute encoder in G-PCC.}
\label{fig:gpcc}
\end{figure}

Besides the GFT-based methods, Queiroz \textit{et al}. proposed a \emph{region-adaptive hierarchical transform} (RAHT) scheme by using the Haar wavelet, which is combined with a Laplacian distributed entropy coder for the attribute compression \cite{RAHT}. This method is also adopted in G-PCC to compress the static point clouds because of its lower complexity. In addition to RAHT, there are another two schemes also for attribute coding, i.e., the \emph{Lifting Transform} method \cite{lift} and the \emph{Predicting Transform} \cite{GPCC} method, which aim for lossy coding and lossless coding respectively. In G-PCC, the Predicting Transform scheme is a distance-based prediction method by using the k-nearest neighbours (KNNs) algorithm. However, this scheme, where the attribute similarity is measured only by the distances between points, fails to fully exploit the geometrical correlation among neighbors.

It is noted that the angle between normals of points can represent the local similarity for LiDAR point clouds. Motivated by this, the optimal mode of the predictor could be more accurately estimated by adding a new descriptor (i.e., normals). In this paper, we propose a normal-based intra prediction scheme for lossless point cloud attribute compression. Based on the original distance-based prediction method, the normals of point clouds are introduced additionally to further optimize predictors, which leads to better prediction mode selections. Extensive experiments are conducted and compared with the original G-PCC, and the experimental results demonstrate that our method provides better R-D performance than the G-PCC anchor for the LiDAR point cloud.

The rest of this paper is organized as follows. Section $2$ describes the lossless attribute compression method in ﻿original MEPG's G-PCC and Section $3$ presents our proposed normal-based intra prediction scheme. Section $4$ provides experimental results and ﻿discussions. Finally, Section $5$ concludes this paper.

\linespread{1.2}
\begin{table}[t]
\small
\begin{center}
\caption{The transform methods for Category $1$ and $3$ under different test conditions in G-PCC.} 
\begin{threeparttable}
\label{tab:cap}
\begin{tabular}{p{1.8cm}p{5cm}p{5cm}} 
\specialrule{0.12em}{3pt}{2pt}
  \multicolumn{1}{c}{\multirow{2}{*}{Dataset}} & \multicolumn{2}{c}{Test condition for attribute} \\ \cline{2-3}
  & \multicolumn{1}{c}{Lossy} & \multicolumn{1}{c}{Near-lossless / Lossless} \\
  \specialrule{0.05em}{1.5pt}{2pt}
  \multicolumn{1}{c}{Category 1} & \multicolumn{1}{c}{RAHT $^{1}$}
  & \multicolumn{1}{c}{Predicting Transform $^{2}$}\\ 
  \multicolumn{1}{c}{Category 3} & \multicolumn{1}{c}{Lifting Transform $^{3}$} & \multicolumn{1}{c}{Predicting Transform $^{2}$}\\ 
\specialrule{0.12em}{2pt}{1.5pt}
\end{tabular}
  \begin{tablenotes}
        \footnotesize
        \item[$1$] region-adaptive hierarchical transform.
        \item[$2$] interpolation-based hierarchical nearest-neighbour prediction.
        \item[$3$] interpolation-based hierarchical nearest-neighbour prediction with an update/lifting step. 
  \end{tablenotes}
\end{threeparttable}
\end{center}
\end{table}

\section{The overview of the attribute encoder in MPEG's G-PCC}
Fig. $2$ shows the framework of the attribute encoder in G-PCC, which mainly consists of the following stages: transferring attributes (the recoloring of reconstructed point clouds), transformation, quantization and entropy coding. In this paper, we work on the transformation stage. 

From Fig. $2$, it can be observed that at the transformation stage, there exists a switch between multiple attribute transform strategies, which depends on various datasets and coding conditions. Specifically, there are two types of datasets provided by MPEG, i.e., the static point clouds (denoted as Category $1$) and the dynamically acquired point clouds (denoted as Category $3$)~\cite{CTC}. As shown in Table $1$, for the lossy coding, the \emph{RAHT} and \emph{Lifting Transform} methods are typically conducted on Category $1$ and Category $3$ respectively. For the near-lossless and lossless coding, the \emph{Predicting Transform} method is used both for Category $1$ and Category $3$.  

\begin{figure*}[htb]
  \centering
  \includegraphics[width=0.95 \textwidth, height=0.46\textheight]{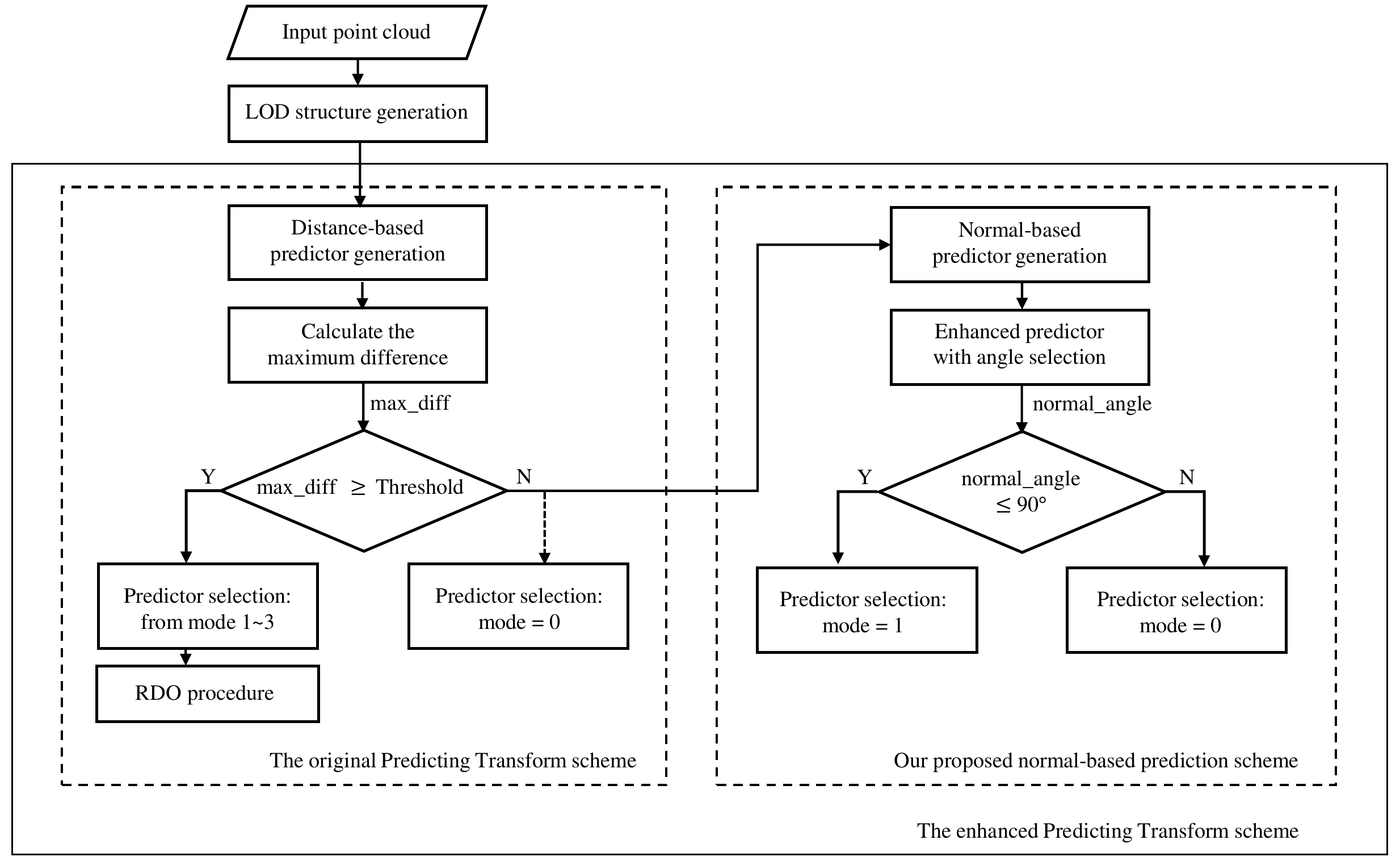}
  \caption{The flow diagram of the enhanced Predicting Transform framework. This scheme consists of two parts: one part is the original Predicting Transform method in G-PCC shown by the dashed box(left), and the other part is our proposed normal-based prediction scheme shown by the dashed box(right). The normal-based prediction consists of three stages, which are the normal-based predictor generation, the predictor enhancement with an angle selection and the predictor modes decision.}
\label{fig:fig3}
\end{figure*}

As previously mentioned, we focus on the attribute lossless coding in G-PCC~\cite{GPCC}, where the \emph{Predicting Transform} method is adopted. This method is an interpolation-based prediction scheme, which is improved by a neighbour weight modification proposed in \cite{pre1} for the LiDAR point clouds. Then, an adaptive predictor selection method~ \cite{pre2} is applied to G-PCC and the rate-distortion optimization (RDO) procedure is used to optimize the prediction scheme, which is modified according to the attribute range of neighbors for LiDAR point clouds, as presented in \cite{pre3}.

\section{The proposed method}
The proposed prediction framework for the lossless attribute compression, as shown in Fig. $3$, consists of the level of detail (LOD) generation and the enhanced Predicting Transform scheme. The LOD structure divides the whole point cloud into a set of refinement levels. Then, the re-organized point cloud is processed by the enhanced Predicting Transform scheme, which incorporates the original Predicting Transform scheme in G-PCC shown by the dashed box (left) and our proposed normal-based prediction scheme shown by the dashed box (right). The Section $3.1.$ introduces the normals' calculation and integration process. The Section $3.2.$ details our proposed normal-based prediction scheme, which introduces the generation, enhancement and selection of the normal-based predictor. 

\subsection{The Integration of the Original Distanced-based Prediction and the Proposed Normal-based Prediction}
In G-PCC, the point cloud is first sorted according to their associated Morton codes in an ascending order, and then divided by the LOD generation process. In our proposed scheme, we preprocess the re-ordered point clouds by calculating the normals before the LOD process, which is prepared for the subsequent normal-based prediction. To be specific, for each point $P_{i}$, we use the k-d tree to find $N$ nearest neighbors (i.e., $N=15$), which defines the local plane of $P_{i}$. Then, the normal of the local plane is calculated by using the eigenvalue decomposition, which serves as the approximate normal of the point $P_{i}$. The point clouds along with calculated normals are next re-organized by the LOD structure process. 

Fig. $3$ presents the enhanced Predicting Transform framework, where the original Predicting Transform scheme is as a part shown by the dashed box (left). It is known that there are mainly three stages in the Predicting Transform scheme: the distance-based predictor generation, the calculation of the attribute value range and the selection of predictor modes. At the stage of the distance-based predictor generation, the distances from the current point $P_{i}$ to previously encoded points are computed and then $k$ ($k=3$) nearest neighbour points of $P_{i}$ can be selected. 

Based on 3 selected points, the distance-based predictor is generated, which is then utilized for the selection of predictor modes. Specifically, the differences between each pair of three nearest neighbors are computed in turn, and then the maximum difference can be available, denoted as \emph{max\_diff}. By comparing the \emph{max\_diff} with a pre-defined threshold, different predictor modes are selected accordingly. In G-PCC, there are four predictor modes, denoted as Mode $0$, Mode $1$, Mode $2$ and Mode $3$ respectively. Mode $0$ represents an interpolation-based prediction with the Inverse Distance Weighted (IDW) method by using $3$ nearest-neighbors. For Mode $1$, Mode $2$ and Mode $3$, $1^{st}$, $2^{nd}$ and $3^{rd}$ nearest neighbors are directly used to predict the current point respectively.

From Fig. $3$, it can be observed that when the \emph{max\_diff} of the neighbor's reflectance exceeds the pre-defined threshold, the current point $P_{i}$ is adaptively predicted from three predictor candidates (Mode $1$, Mode $2$ and Mode $3$) by using the rate-distortion optimization (RDO) procedure. Otherwise, Mode $0$ will be chosen. In this case, instead of using Mode $0$, our proposed normal-based predictor is applied, which aims to optimize the selection of predictor candidates. The specific procedure is shown in Fig. $3$ denoted with the dashed box (right). The main novelty is that, in addition to the predictor in the original G-PCC using the similarity in distances between coordinates, our proposed method further explores the local similarity among points by introducing the angle between normals.

\linespread{1.2}
\begin{table*}[htb]
\small
\begin{center}
\caption{The comparison results between our method and the G-PCC anchor on Category 3-frame under the condition CW.} 
\label{result}
\begin{tabular}{p{4cm}p{1.8cm}p{2cm}p{2cm}p{1.8cm}p{2.4cm}} 
\specialrule{0.12em}{5pt}{2pt}
  \multirow{2}{*}{Sequences} & \multicolumn{1}{c}{Frame}  & \multicolumn{1}{c}{\multirow{2}{*}{Input Points}} & \multicolumn{1}{c}{Geometry} & \multicolumn{1}{c}{Peak} & \multicolumn{1}{c}{\multirow{2}{*}{$\Delta R$ (\%)}}
  \\ & \multicolumn{1}{c}{Number} & &\multicolumn{1}{c}{Precision (bits)}& \multicolumn{1}{c}{Value} & \\
  \specialrule{0.05em}{1.5pt}{2pt} 
  {ford\_01\_q1mm} & \multicolumn{1}{c}{1500} & \multicolumn{1}{c}{123940658} & \multicolumn{1}{c}{18} & \multicolumn{1}{c}{30000} & \multicolumn{1}{c}{-1.3}\\ 
  {ford\_02\_q1mm} & \multicolumn{1}{c}{1500} & \multicolumn{1}{c}{125751705} & \multicolumn{1}{c}{18} & \multicolumn{1}{c}{30000} & \multicolumn{1}{c}{-1.0}\\ 
  {ford\_03\_q1mm} & \multicolumn{1}{c}{1500} & \multicolumn{1}{c}{126093865} & \multicolumn{1}{c}{18} & \multicolumn{1}{c}{30000} & \multicolumn{1}{c}{-0.9}\\ 
  {qnxadas-junction-approach} & \multicolumn{1}{c}{74} & \multicolumn{1}{c}{2233793} & \multicolumn{1}{c}{18} & \multicolumn{1}{c}{30000}  & \multicolumn{1}{c}{-2.8}\\
  {qnxadas-junction-exit} & \multicolumn{1}{c}{74} & \multicolumn{1}{c}{2016190} & \multicolumn{1}{c}{18} & \multicolumn{1}{c}{30000}      & \multicolumn{1}{c}{-5.6}\\ 
  {qnxadas-motorway-join	} & \multicolumn{1}{c}{500} & \multicolumn{1}{c}{14430189} & \multicolumn{1}{c}{18} & \multicolumn{1}{c}{30000} 	  & \multicolumn{1}{c}{-4.5}\\ 
  {qnxadas-navigating-bends} & \multicolumn{1}{c}{300} & \multicolumn{1}{c}{8167066} & \multicolumn{1}{c}{18} & \multicolumn{1}{c}{30000}   & \multicolumn{1}{c}{-4.0}\\ 
  \specialrule{0.05em}{1.5pt}{2pt}
  {Cat3-frame average} & & & & & \multicolumn{1}{c}{-2.1}\\ 
  \specialrule{0.12em}{2pt}{1.5pt}
\end{tabular}
\end{center}
\end{table*}

\subsection{The Generation, Enhancement and Selection of the Normal-based Predictor}
The normal-based prediction method is shown in Fig. $3$ with the dashed box (right), which mainly consists of three stages: the normal-based predictor generation, the predictor enhancement with an angle selection and the predictor modes decision. To be specific, the normal-based predictor is generated on top of the previous distance-based predictor, by introducing the normals of $3$ nearest-neighbors. 

The normal-based predictor obtained above is then improved by an angle selection. The calculation of the angle between normals is shown in Fig. $4$. Let $V_{i}$ and $V_{j}$ be the normals of the points $P_{i}$ and $P_{j}$ respectively. Then, the angle $\theta$ between normals $V_{i}$ and $V_{j}$ is calculated by
\begin{eqnarray}
\theta &=& \arccos \left ( \frac{V_{i} \cdot V_{j} }{\left \|V_{i} \right \|\cdot \left \|V_{j} \right \|} \right ). \label{eq4}
\end{eqnarray}
At the stage of the predictor enhancement, considering the similarity of the attributes decreases with the increase of the distance between point clouds, we only use the $1^{st}$ nearest neighbouring point, which is to enhance the normal-based predictor with an angle selection. Specifically, we calculate the angle between normals of $1^{st}$ nearest neighbor and the current point based on Equation \eqref{eq4}, denoted as the $normal\_angle$.

From Fig. $3$, it can be observed that when the $normal\_angle$ is greater than $90^{\circ}$, Mode $0$ is selected by using the weighted average of neighbors' attributes for the prediction. If not, Mode $1$ is selected by using the $1^{st}$ neighbor to predict the current point. Note that Mode $0$ and Mode $1$ follow the same paradigm adopted in G-PCC.

\begin{figure}[htb]
  \centering
  \includegraphics[width=0.40\textwidth, height=0.2\textheight]{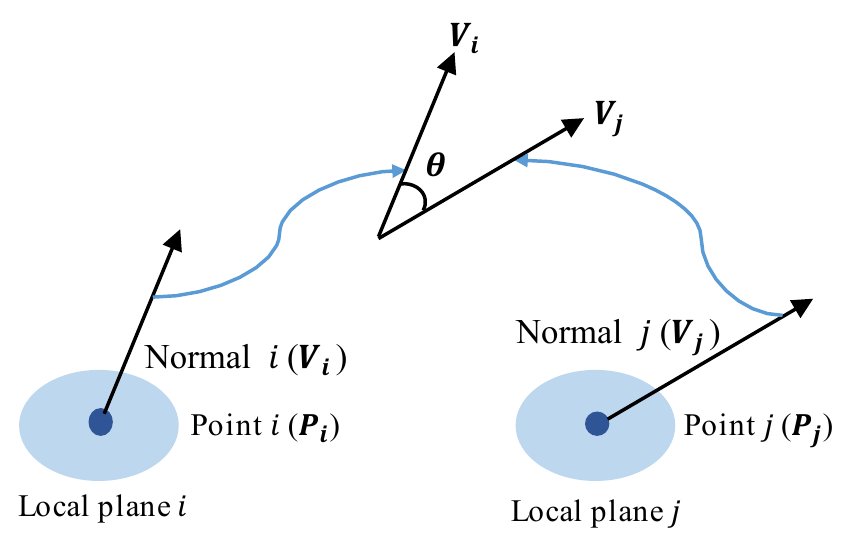}
  \caption{The schematic diagram of calculation for the angle between normals. $V_{i}$ and $V_{j}$ are normals of point $P_{i}$ and $P_{j}$ respectively. The angle between normals of points is denoted as $\theta$.}
\label{fig:angle}
\end{figure} 

Finally, the prediction residuals are processed by the quantizer and arithmetic encoder. It is worth to mention that no extra flags or parameters are required to be written into the final bitstream, because the decoder is implemented by repeating the operation of the normal-based prediction scheme. 

\section{experimental results}
To evaluate the effectiveness of our proposed normal-based intra prediction scheme, extensive simulations have been conducted on the test dataset provided by MPEG. We incorporate our method into the MPEG G-PCC reference software, i.e., TMC13v10~\cite{GPCC}, and compare its compression performance~\cite{pce} of attribute compression with the original TMC13. 

In our experiments, the test dataset consists of seven dynamically acquired point clouds (denoted as Category $3$-frame). Among them, three ford sequences, \emph{ford\_01\_q1mm}, \emph{ford\_02\_q1mm}, and \emph{ford\_03\_q1mm} can be available by \cite{ford}, while the other four qnxadas sequences, \emph{qnxadas-junction-approach}, \emph{qnxadas-junction-exit}, \emph{qnxadas-motorway-join}, and \emph{qnxadas-navigating-bends}, can be obtained in \cite{qnx}. More details of the Category$3$-frame dataset are listed in Table~\ref{result}. All the experiments are conducted under CW condition of the Common Test Conditions (CTC)~\cite{CTC}, where CW represents lossless geometry and lossless attribute. 

Table~\ref{result} shows the comparison results between proposed method and the G-PCC anchor on the Category $3$-frame under the CW condition. Since this work aims for lossless compression with no distortion of data, the commonly-used evaluation metric, measuring the rate in terms of bits for attributes, bits per input point (denoted as $bpip$), is used for evaluation. Specifically, we compute the $bpip$ of the proposed method and that of the TMC13, and then the bit saving ratio ($\Delta R$) can be obtained for the final performance evaluation, which is defined as

\begin{equation}
\Delta R=\frac{bpip_{ours}-bpip_{tmc13}}{bpip_{tmc13}} \times 100 \% \label{ratio}
\end{equation}

From equation~\eqref{ratio}, we can see that the proposed method outperforms the anchor (TMC13) when the bit saving rate $\Delta R$ is negative. Otherwise, it means that the anchor shows better performance than the proposed method.

From Table~\ref{result}, it can be obviously observed that the bit savings can be achieved with 2.1\% on average, especially for sequence \emph{qnxadas-junction-exit}, up to 5.6\%. From the results, we can see that our proposed method is effective on the LiDAR datasets for attribute lossless compression in G-PCC. The rationale behind the gains is that by introducing our normal-based prediction scheme, the attribute similarity can be explored more accurately, which leads to better selection of the predictor mode.

\section{conclusion}
In this paper, to improve the efficiency of the lossless attribute compression in G-PCC, a normal-based intra prediction scheme is proposed, which further exploits the geometrical correlations among neighbors in point clouds. Based on the original distance-based Predicting Transform scheme, the normals of each point, as an additional descriptor, are introduced to optimize original predictors. By computing the angle between normals, a better predictor mode can be selected. Experimental results have demonstrated that our method is able to consistently deliver a better performance than the G-PCC test model.

\bibliographystyle{IEEEbib}
\bibliography{reference}

\end{document}